\documentclass{article}
\usepackage{graphicx} 
\usepackage{amsmath}
\usepackage{amsfonts}
\usepackage{rsfso}
\usepackage{hyperref}
\usepackage[
backend=biber,
style=numeric,
sorting=none,
sortcites=true
]{biblatex}

\title{The Ergodic Vacuum}
\author{Chris Scott \footnote{Advanced Energetics Research Group, Royal Australian Air Force} }
\date{May 2024}

\begin{document}

\maketitle

\begin{abstract}
    The extension of local de Sitter thermodynamics into $f(\mathcal{R})$ gravity provides a new basis to unify the dielectric and electroweak vacua. We suggest the electroweak theory emerges from the ergodic mixing of charge density under local de Sitter thermodynamics and introduce the concept of lab-accessible topological defects.  
\end{abstract}
It is not possible for the Ward identity to fix the vacuum polarization tensor, contrary to (5.79) of \cite{peskin_introduction_2019}, for the following reason: 
\begin{itemize}
    \item Inherent in the Ward-Takahashi identity is the Siegert theorem.\cite{friar_gauge_1992} 
    \item The Siegert theorem is invalid for toroid polarized media.\cite{dubovik_toroid_1990}
    \item The vacuum polarization tensor of \cite{peskin_introduction_2019} is subjected to an artificial gauge choice if the media's polarization tensor necessarily cannot contain a specific polarization. 
 \end{itemize}
This is the current Field Theory dilemma which we seek to rectify through an applied condensed matter approach.\\

Local de Sitter thermodynamics relates the scalar radius $\mathcal{R} = -12R^2$ to the de Sitter scaling factor $R$ via a toroidal connection with the Riemann curvature \cite{volovik_sitter_2024}, in flat cartesian space,
\begin{align} 
    R_{\alpha\beta\mu\nu} = (\epsilon_{\alpha\beta\lambda}\epsilon_{\mu\nu\lambda})R^2
\end{align}
Similarly the FLRW metric can be represented by the toroidal de Sitter metric with a singularity horizon at radius $R$.\cite{numasawa_global_2019} 
\begin{align} \label{eq:ab}
     \quad g_{\alpha\beta} = \text{diag}\big({e^{Rt}r_x, e^{Rt}r_y, e^{Rt}r_z, \frac{1}{R}\cosh{Rt} - \frac{Rr^2}{2}e^{Rt}, \frac{1}{R}\sinh{Rt} + \frac{Rr^2}{2}e^{Rt}}\big)
\end{align}
 
Running scale in a fluid picture, momentum flux terminates at the wavenumber $k_{max} = {r_{0}^{-1}}$ where $r_{0}$ is the minimum accessible radius of the path of a charged, massive particle. Classic treatments such as \cite{kolmogorov_local_1991} replace momentum flux with Coulomb interactions up to $\mathcal{O}(2)$, consistent with the velocity field perturbations. In contrast, accounting for the relative rotation of eddy ensembles in a fluid requires $\mathcal{O}(3)$ terms of the velocity (and Coulomb) fields. For this same reason the dual rotation of both EM sources \textit{and} fields, termed \textit{dyality}, gives rise to an extra $U(1)$ symmetry.\cite{han_manifest_1971, dubovik_material_2000}  

At the terminal energy cascade scale we might view a super-domain tiled with oscillator domains similar to a spin-glass picture except in this case the lattice is unconstrained. As the conjugate angular momentum approaches the terminal scale, $\sim r_{0} $, we consider the dynamics of the non-dissipative oscillator as being \textit{in involution}. For such a system with $n$ degrees of freedom the phase space is free to explore $2n$ degrees of freedom and the corresponding energy surface has ($2n-1$) dimensions. This phase space is a manifold described by the $n$-torus.\cite{berry_regular_1978} 

\begin{figure}
    \centering
    \includegraphics[width=1\linewidth]{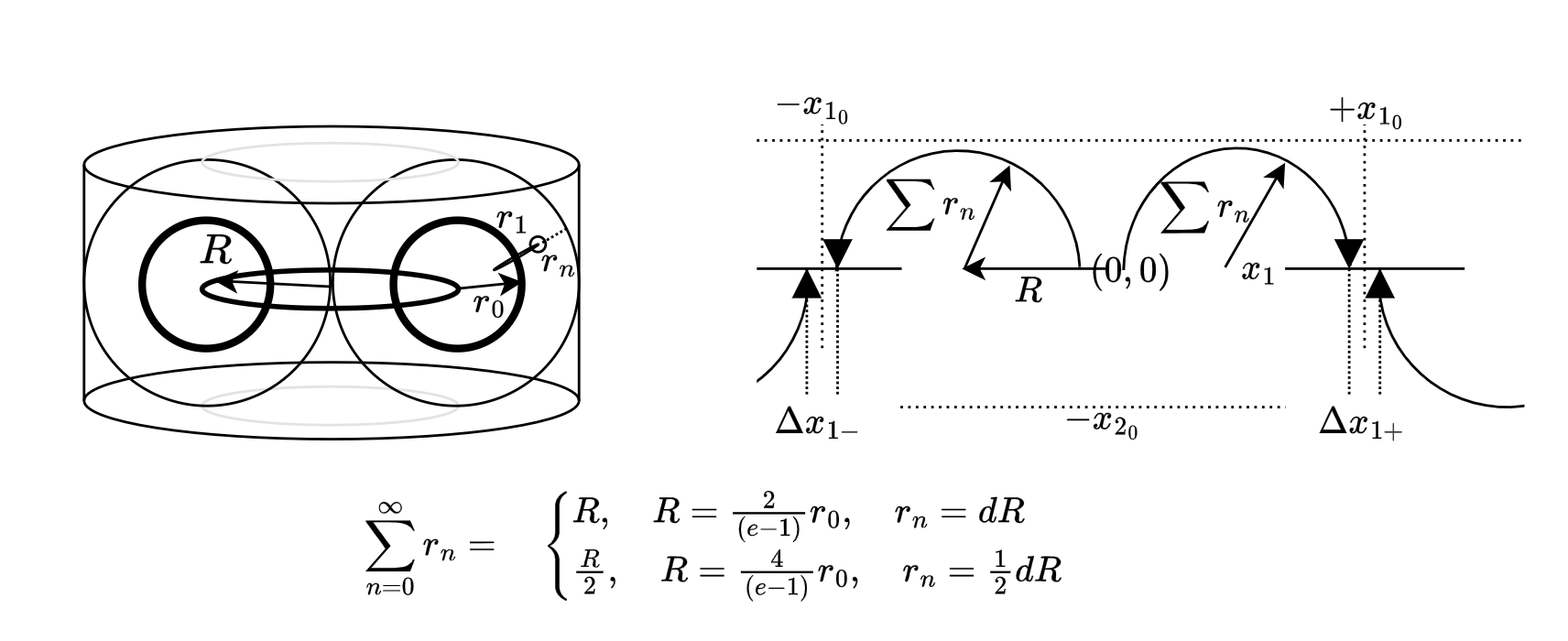}
    \caption{Hypertorus geometry and the relationship between the singularity horizon at $R$ within the cylindrical, spin-cell domain. The solution and vacuum polarization spectrum were calculated for the case $r_n = dR$, however, for the case $r_n = \frac{1}{2}dR$, the solution holds with the familiar normalization, $\frac{1}{2}\sigma$.}
    \label{fig:hypertorus_geo}
\end{figure}

\begin{figure}
    \centering
    \includegraphics[width=1\linewidth]{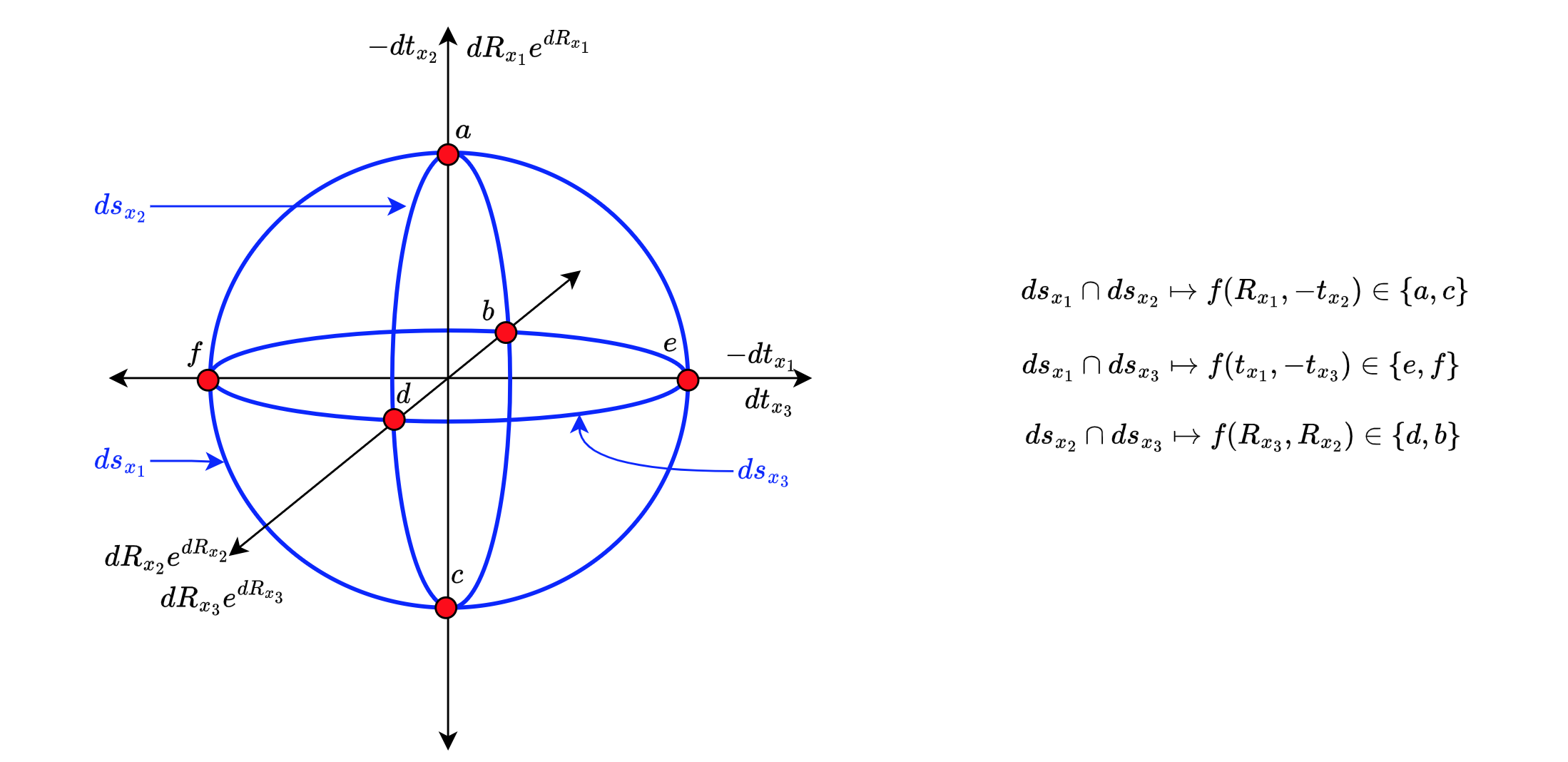}
    \caption{The ergodic solution space is given by the six intersections of three orthogonal de Sitter paths when endowed with $r_n = dR$, see fig \ref{fig:hypertorus_geo}. The choice of two intersections fixes four dimensions of the $\dim 5$ energy surface and corresponds to the symmetry breaking of the $SU(2)$ symmetry group inherent in this geometric representation.}
    \label{fig:soln_space}
\end{figure}

Since the $n$-torus is orientable, and because the viscosity is defined on a 2d plane only,\footnote{The rate of strain tensor has the same form factor as a quadrupole} the intersection of three $n$-tori define the domain of solutions to both the Hamiltonian and the change in scalar curvature $\mathcal{R}$ as applied to the viscosity dominated fluid. The infinitesimal total electric flux displacement is the toroidal de Sitter path on the $n$-th hypertorus symmetric about $x_i$ which has the notation $(ds_n)_{x_i}$. The enveloping toroid moment is given by the usual integral, 
\begin{align} \label{eq:1onten}
    \vec{T}_i^{n(m)} = \Big( \frac{1}{10c} \Big)^{n} \int (\vec{r}\vec{r}\cdot\vec{T}^{(n-1)(m)} - 2r^2\vec{T}^{(n-1)(m)})d^3r
\end{align}
\begin{align}
    \vec{T}^{0(m)} =  \int (\vec{r}\vec{r}\cdot \phi^t - 2r^2 \phi^t )d^3r
\end{align}
where $\phi^t$ is the total current vector and $\vec{r} = R + \frac{2}{(e-1)}n!r_0$ at $x_j=x_{j_0}-\Delta x_j /2$, fig (\ref{fig:hypertorus_geo}). The non-dissipative orbit of the massive particle doesn't leave the n-torus so the total flux contribution to (\ref{eq:1onten}) is invariant so long as the $n$-torus and integration domain of (\ref{eq:1onten}) share the same orientation. The contribution of the neighbouring spin domains to (\ref{eq:1onten}) is encapsulated in an octupole moment in the plane $ij$, the $\mathbf{irrep}$ of the toroid moment. This is the viscous contribution being the two rate of strain tensors centred on opposing sides at $\pm x_{1_0}$ in fig (\ref{fig:hypertorus_geo}). Integrating only outside the non-dissipative torus with minor radius $r_0$ results in the following hypergeometric function. 
\begin{align} \label{eq:Tn}
  \vec{T}^{n(m)}_i = -\frac{2}{3}\frac{1}{10c}\Big( \frac{2r_0}{(e-1)}n!\Big)^3 \Bigg( \frac{\sum r_n}{2\sum r_{(n-1)}}\Bigg) \Big( \frac{1}{4 \pi c}\Big)^{(n-1)}
\end{align}
Imposing rotation invariance on (\ref{eq:Tn}) gives solutions for $n$, the toroidal hypersurface on which the normalised electric field flux $\vec{E}$ path is confined. The change in $n$ only comes from the external contributions over both $ \Delta x_j$ domains of the octupole, i.e. shear contributions. Since both orientation of the toroid dipole moment and $\Delta n \sim \ dR$ (fig \ref{fig:hypertorus_geo}) are the only degrees of freedom the solutions of (\ref{eq:Tn}) correspond to possible toroidal de Sitter orientations and change of scale factor $R$. The $n$-th radius $r_n$ contributes to the change in $R$ according to the spacing of the spin cells within the super-domain. 
\begin{align} \label{eq:space_cases}
     \sum_{n=0}^{\infty} r_n =  \quad \begin{cases}  R, \quad R = \frac{2}{(e-1)}r_0, \quad r_n = dR, \quad \Delta x \ll R \\  \frac{R}{2}, \quad R = \frac{4}{(e-1)}r_0, \quad r_n = \frac{1}{2}dR, \quad \Delta x = R \end{cases}
\end{align}
It's straightforwad to show the first condition in (\ref{eq:space_cases}) maximises viscosity. With this substitution the de Sitter path becomes,
\begin{align} \label{eq:desitterpath}
    ds^2_{x_i} = dt^2_{x_i} + (dR_{x_i})^2e^{2dR_{x_i}}
\end{align}
So that the intersections in (\ref{eq:intesctn}) can be written in terms of $R_i$ only for three orthogonal toroidal de Sitter spaces. (See fig (\ref{fig:soln_space}).)
\begin{align} \label{eq:intesctn}
    (ds_n)_{x_1} \cap (ds_n)_{x_2} =
    (ds_n)_{x_2} \cap (ds_n)_{x_3} = 
    (ds_n)_{x_3} \cap (ds_n)_{x_1}
\end{align} The obvious $U(1)$ symmetry in (\ref{eq:desitterpath}) has the infintesimal components $dt_{x_i}$ and $dR_{x_i}\exp({dR_{x_i}})$ where $dt_{x_i}$ is defined on closed paths symmetric about $\hat{x_i}$. Fig \ref{fig:soln_space} shows the local turbulence solution space when energy-minimised, all-order EM interactions determine both the \textit{orientation} and \textit{polarization} states. \cite{scott_turbulence_2024} 

This picture closes the Sommerfeld puzzle to the extent that both Hamiltonian and Lagrangian are provided equal footing.\footnote{An equivalent Proca field approach would require relaxed constraints due to their decomposition under the dyality U(1) symmetry. The Vorton model achieves this decomposition and \cite{garaud_stable_2013} additionally invokes the radial gauge, applicable to the local de Sitter thermodynamic model. } The available polarization states are defined by the solutions of the PDEs in (\ref{eq:sigma_munu}) and (\ref{eq:sigmas}) given the usual Pauli matrices, $\sigma$ embedded into the Minkowski metric. The effect is twofold; a) third order perturbations are effectuated under contraction of $\sigma'_{\mu\nu}$ with the usual EM tensor, and b) the effective compactification is achieved via,
\begin{align} \label{eq:sigma_munu}
    \sigma'_{\mu\nu} = \text{diag}(1,1,1,\sigma)
\end{align}
 \begin{align} \label{eq:sigmas}
     \sigma = \begin{pmatrix}
       \frac{\partial}{\partial t_{\parallel}}\frac{\partial}{\partial t_{\parallel}}  & \frac{\partial}{\partial t_{\parallel}}\frac{\partial}{\partial t_{\perp}}\\
        \frac{\partial}{\partial t_{\perp}}\frac{\partial}{\partial t_{\parallel}} & \frac{\partial}{\partial t_{\perp}}\frac{\partial}{\partial t_{\perp}}\\
    \end{pmatrix} = \begin{cases}
    \sigma_1 = \begin{pmatrix}
        0 & 1\\
        1 & 0\\
    \end{pmatrix} \\
    \sigma_2 = \begin{pmatrix}
        0 & -i\\
        i & 0\\
    \end{pmatrix} \\
    \sigma_3 = \begin{pmatrix}
        1 & 0\\
        0 & -1\\
    \end{pmatrix} \\
\end{cases}
 \end{align}
The geometrical interpretation of (\ref{eq:sigma_munu}) is a three sphere with radius $\sigma$. The $SU(2)$ symmetry of the Pauli matrices act \textit{transitively} on $S^3$ so that the dilaton, $\frac{\partial^2 X}{\partial t_{\parallel}^2}, X(x_1,x_2,x_3) \in \mathbb{R}^3$, and chirality take the same sign of the compact temporal dimension. This compact dimension identifies the Kaluza-Klein mysterious 5th dimension specifically arising under the relative rotation (acceleration). The applicable transformation matrix $M^{\alpha\beta}$ corresponds to the $\dim(2n-1)$ energy surface where the spatial and temporal components together represent electric multipoles of $\mathcal{O}(2)$ and $\mathcal{O}(3)$ respectively. This is simply the decomposition of the usual EM four-potential $A=(\phi,\vec{A})$ and its dyality dual $\Pi = (\phi^{(m)}, \vec{T})$ into compact and transverse temporal components in (\ref{eq:ab}),
\begin{align} \label{eq:M}
    M^{\alpha\beta} = \partial^{\alpha} A^{\beta} - \partial^{\beta}A^{\alpha} - 4\pi (\partial^{\alpha} \Pi^{\beta} - \partial^{\beta} \Pi^{\alpha})
\end{align}

$M^{\alpha\beta}$ is the ergodic $\dim 5$ transformation matrix which conserves the total electric flux displacement in the net zero magnetic flux condition. In this condition as the penultimate angular momentum cascades into the highest accessible wave-number, viscous effects dominate and the local magnetic flux is cancelled. Setting $m = c= 1 $, the \textit{gridlocked} picture ensues when massive, charged particles transfer their momentum via Coulomb interactions only, being stationary with respect to the distance to their nearest neighbour. In this definition an ensemble of particles may be gridlocked but the ensemble subject to relative rotation. In this case we have a radiation dominated fluid for which the toroidal de Sitter flat space is appropriate.

\begin{figure}
    \centering
    \includegraphics[width=1\linewidth]{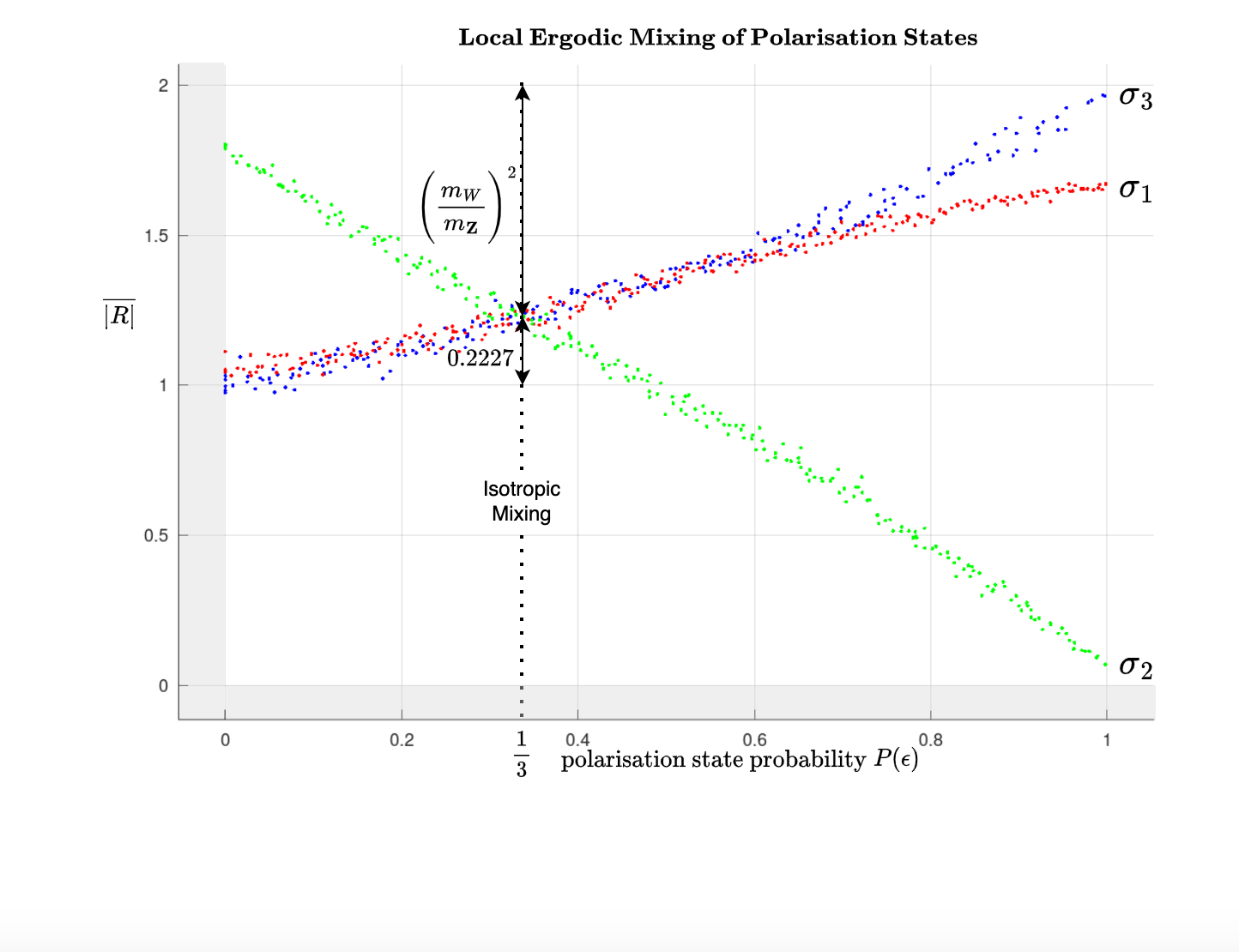}
    \caption{Polarization spectrum where $P(\epsilon)$ is the probability weighting of polarizations $\{ P(\epsilon), \frac{1- P(\epsilon)}{2}, \frac{1- P(\epsilon)}{2} \}.$ Unsurprisingly the weak mixing angle is emergent at the isotropic polarization mixing. Each dot represents the mean absolute scaling factor of the ergodic evolution over 1000 time-steps.}
    \label{fig:spin_cell}
\end{figure}

\begin{figure}
    \centering
    \includegraphics[width=1\linewidth]{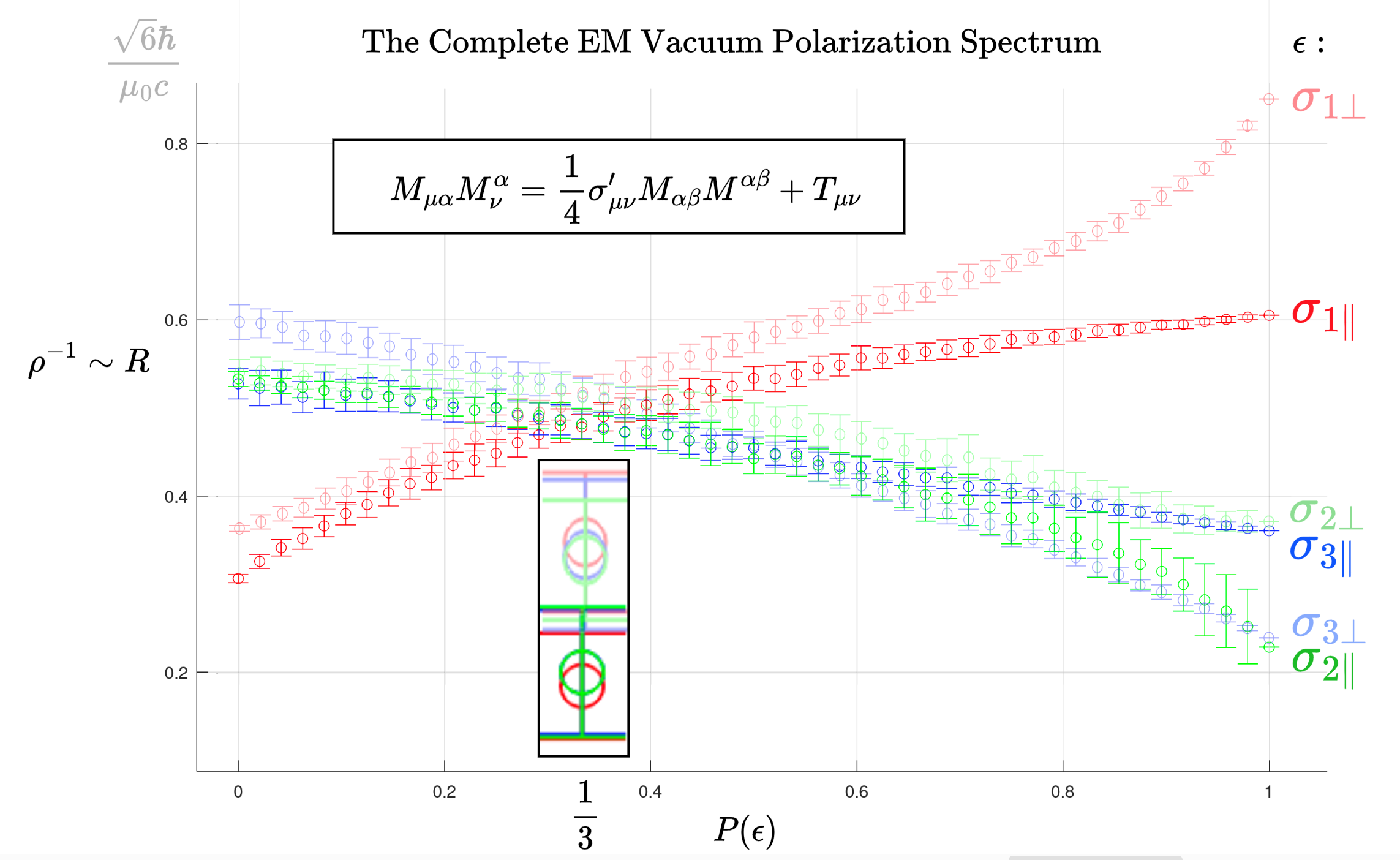}
    \caption{Transverse and compact (longitudinal) polarization spectrum 
 over the range of de Sitter thermodynamic states. $T_{\mu\nu}$ is initialised at a vacuum contribution, i.e. zero. At an isotropic mixing state the Coulomb gauge appears emergent at low EM densities. Each range represents the distribution of the mean absolute scaling parameter $R$ over 50 runs, each run produced a mean value from 1000 time-steps, bars indicate $\pm$ one standard deviation. }
    \label{fig:complete}
\end{figure}

\begin{figure}
    \centering
    \includegraphics[width=0.75\linewidth]{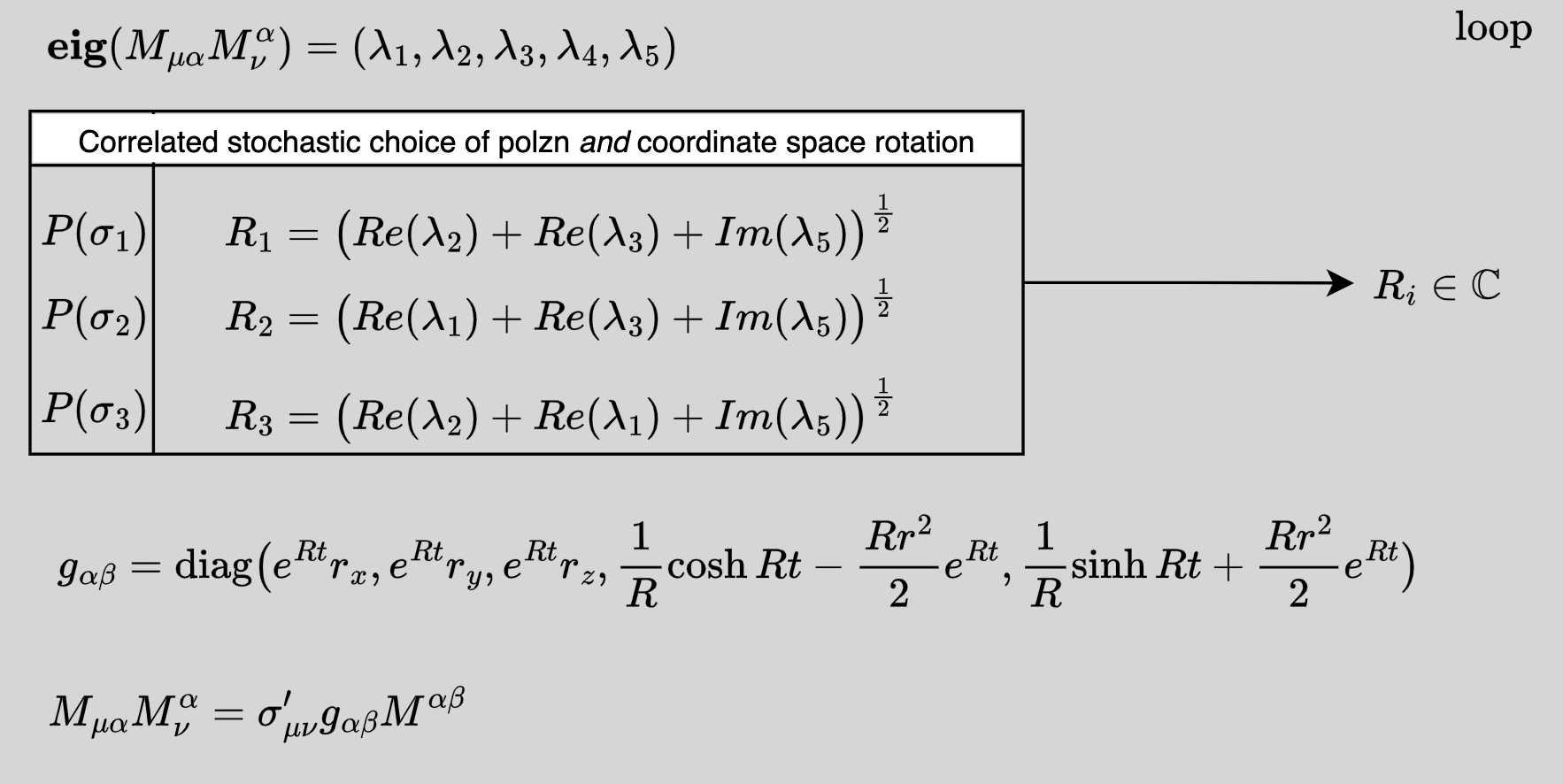}
    \caption{Numerical scheme for the time-evolution of the scale factor $R_i$. Permutation of scaling factor occurs in two dimensions where the orientation and polarization are approximated via a stochastic choice. The second step is equivalent to spontaneous symmetry breaking insofar as the roots of the cubic are solved, in this case in the $x_4$, i.e. compact temporal component normal to either, $x_1, x_2,$ or $x_3$. Imaginary components contribute to the system's available degrees of freedom. This regime was used to solve the discrete time steps in fig \ref{fig:noise} and ultimately fig \ref{fig:spin_cell}.  }
    \label{fig:numerical}
\end{figure}

\begin{figure}
    \centering
    \includegraphics[width=1\linewidth]{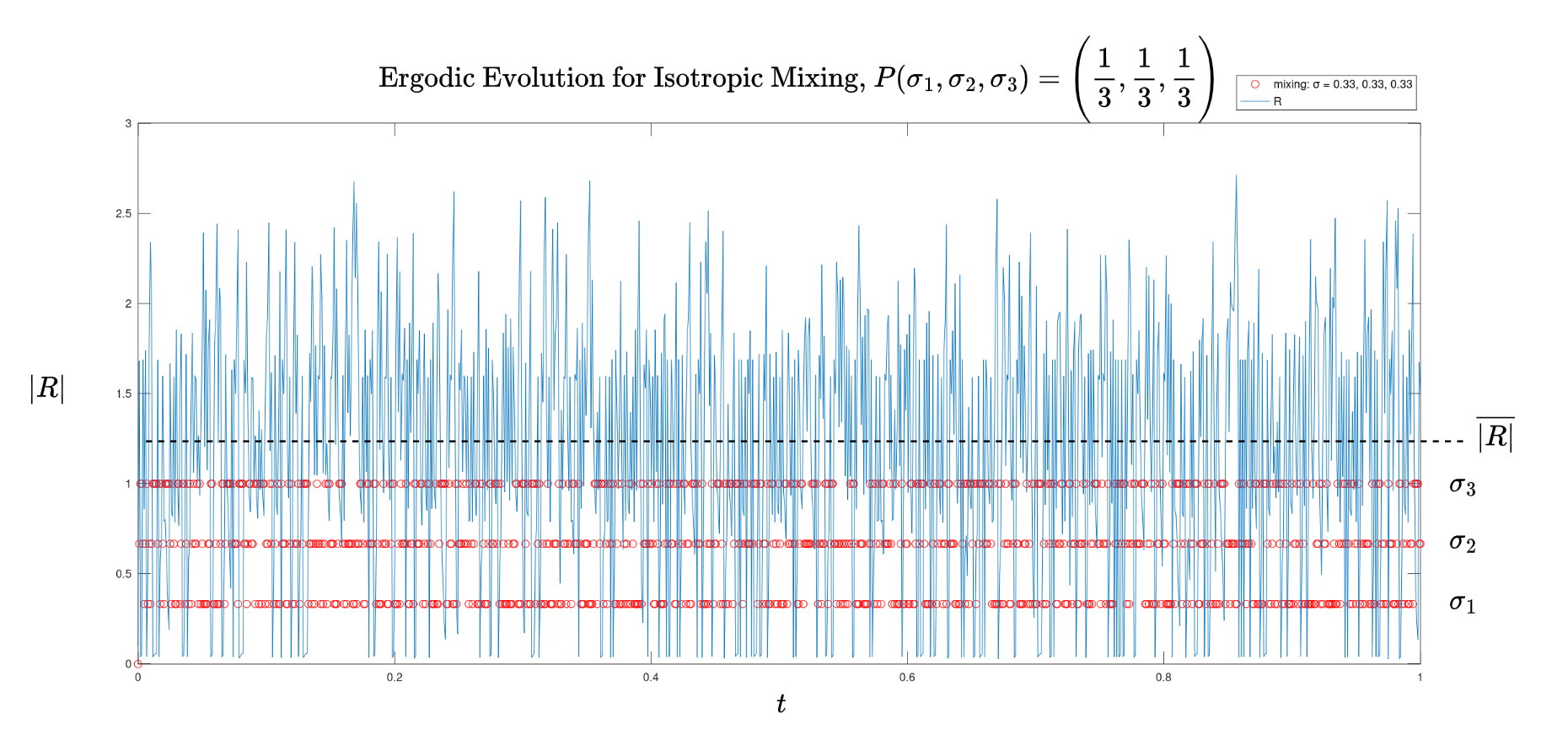}
    \caption{Absolute value of scale factor $R$ (blue line) over 1000 time steps results in a single mean value for a simulation run (black dashed line). The stochastic choice of polarization for discrete time steps is indicated with the red circle. Equal probability of polarization choice simulates the isotropic vacuum fluctuations in the compact domain, annotated in fig \ref{fig:spin_cell}.}
    \label{fig:noise}
\end{figure}

The scale parameter $R$ can be solved numerically for a stochastic choice of polarization $\sigma \in \{\sigma_1,\sigma_2,\sigma_3\}$. (The numerical scheme is in Fig \ref{fig:numerical} and results in Fig \ref{fig:noise} and Fig \ref{fig:spin_cell}) The Kaluza-Klein solution assumes an intrinsic contribution to the stress energy tensor, $T_{\mu\nu}$ applicable to the Einstein-Klein-Gordon equations.\cite{brihaye_kaluza-klein_2024} In this case, 
\begin{align}
    g_{\mu\nu} \simeq \frac{M_{\mu\alpha}M^{\alpha}_{\nu}}{M_{\alpha\beta}M^{\alpha\beta}}, \quad M_{\mu\alpha}M^{\alpha}_{\nu} = \sigma'_{\mu\nu}g_{\alpha\beta}M^{\alpha\beta}
\end{align}
where $g_{\alpha\beta}$ is the ergodic metric (\ref{eq:ab}) applicable to the spin domain.
The general polarization spectrum applicable to the EM vacuum can be numerically solved independent of the Kaluza-Klein model, however, allowing for spectral dispersion mediated by the stress energy tensor (the transverse ergodic component), Fig \ref{fig:complete} with relation inset.  

\begin{figure}
    \centering
    \includegraphics[width=1\linewidth]{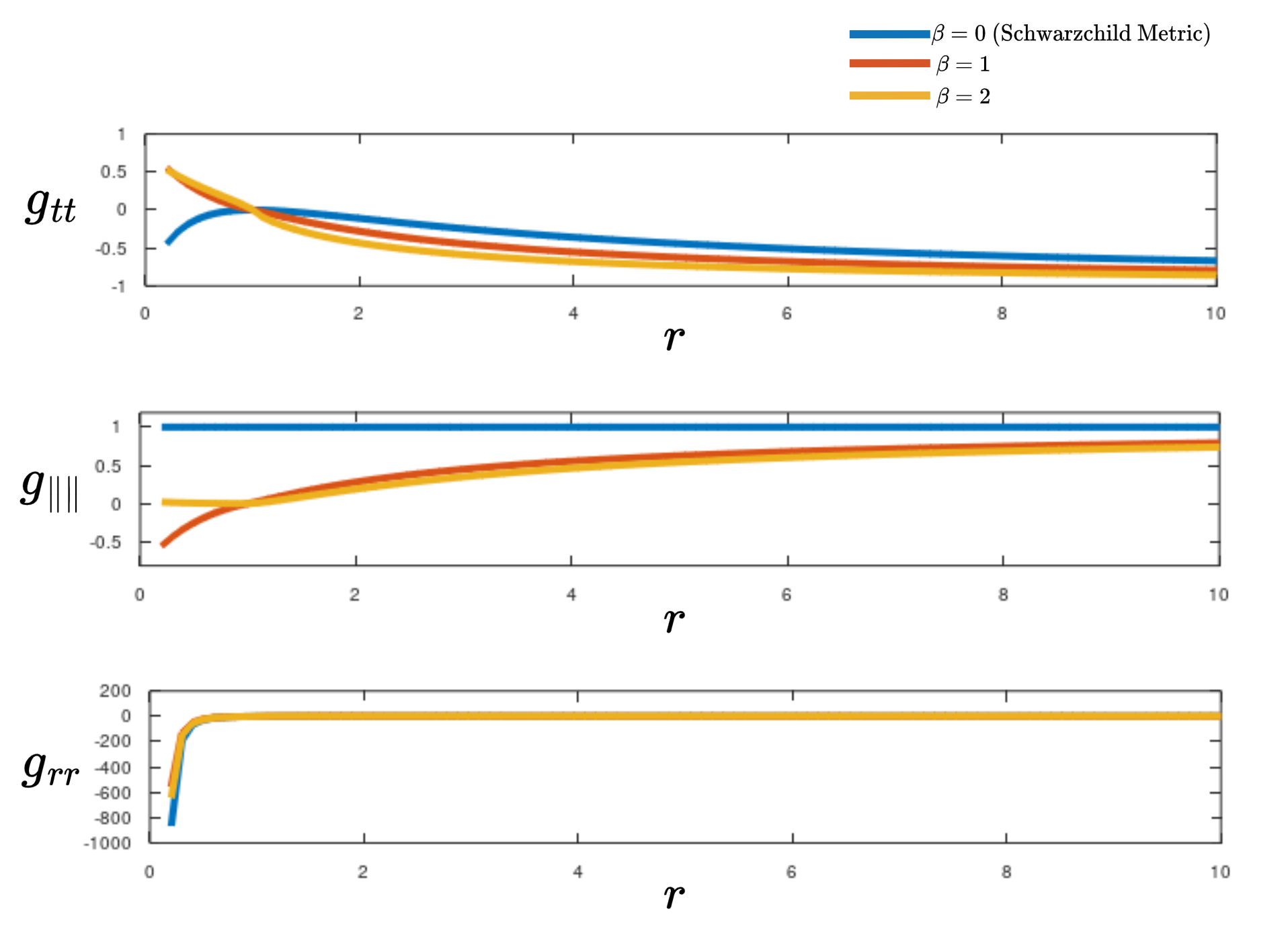}
    \caption{Kaluza-Klein metric solutions for eq \ref{eq:dslong} with varying $\beta$ from \cite{gross_magnetic_1983}. Soliton solutions ($\beta \gg 1$) have a singularity horizon at $r=R=1$ and exhibit inertial mass yet no gravitational mass. $g_{\parallel\parallel}$ is the compact dimension, $g_{44}$ in our numerical scheme, Fig \ref{fig:numerical}.}
    \label{fig:KK_metric_soln}
\end{figure}

Fig \ref{fig:complete} suggests transverse electromagnetic wave solutions are decoupled from the longitudinal (compact) propagating solutions due to the minimal overlap in scalar radius $\sim R$ at isotropic mixing. To properly grasp the dynamics of the compact solutions we turn to the Kaluza-Klein interpretation. There is a whole family of two-parameter, spherically symmetric topological solutions shown in Fig \ref{fig:KK_metric_soln} given by,
\begin{multline} \label{eq:dslong}
    (ds)^2 = -\left(\frac{1 - \frac{m}{r}}{1 + \frac{m}{r}}\right)^{\frac{2}{\alpha}} (dt)^2 + \left(1 + \frac{m}{r}\right)^4 \left(\frac{1 - \frac{m}{r}}{1 + \frac{m}{r}}\right)^{\frac{2(\alpha - \beta - 1)}{\alpha}} (dr^2 + r^2 d\Omega^2) \\+ \left(\frac{1 - \frac{m}{r}}{1 + \frac{m}{r}}\right)^{\frac{2\beta}{\alpha}} (dx_{\parallel})^2
\end{multline}
where $x_{\parallel}$ is periodic and $\alpha = \sqrt{\beta^2 + \beta + 1}$. Regular black hole geodesics emerge for $\beta = 0$ but the soliton solutions ($\beta \gg 1$) exhibit more exotic behaviour; they possess inertial mass but zero gravitational mass. \cite{gross_magnetic_1983} We suggest that hypercharge screening can occur at significantly lower energy density in the topological soliton than e.g. magnetically charged black holes \cite{maldacena_comments_2021} or vortons.\cite{garaud_stable_2013}  The condensation of W bosons as near-field EM excitations of a rotating fermionic condensate may arise under the effective polarization ($F^2 = F^3, F^1 \rightarrow 0$) as a corollary of \cite{ambjorn_electroweak_1993}. The weak coupling constant $g$ may be (anti)screened due to the dyality rotation of EM from which SU(2) emerges. Returning to the Weinberg picture,
\begin{align} \label{eq:F}
    F_{\mu\nu}^a = \partial_\mu W_\nu^a - \partial_\nu W_\mu^a + g \epsilon^{abc} W_\mu^b W_\nu^c
\end{align}
If we presume to have solved the Sommerfeld puzzle, comparing (\ref{eq:M}) and (\ref{eq:F}) gives the compactification,
\begin{align}
    M_{\alpha\beta}\frac{1}{2}\sigma'_{\mu\nu} = F_{\mu\nu}^a
\end{align}
Where the factor $\frac{1}{2}$ arises from the case $R = \frac{4}{(e-1)}r_0$  in Fig \ref{fig:hypertorus_geo}. Extending the relation in the cylindrical domain forms a triangular lattice structure with $3R$ separation. This suggest the electroweak theory emerges when the spin glass model has a triangular lattice arrangement where $R$ is the degree of freedom in (\ref{eq:intesctn}). 

The Kaluza-Klein theory successfully unifies gravity and electromagnetism and paves the way towards solving the cosmological constant problem.\cite{klinkhamer_propagating_2017} Callan-Rubikov candidate phenomena may appear in high-shear, high impulse thermodynamic states\cite{mk_mechanical_2018, sobolev_shock-wave_2006} rather than equilibrium baths.\cite{sreekantan_searches_1984} 
We showed the artificial gauge choice of the Ward identity is only adequate for an isotropic vacuum. The full spectrum of non-equilibrium thermodynamics requires an unambiguous treatment of the vacuum polarization tensor. Local de Sitter thermodynamics recovers the electroweak vacuum properties when the complete multipole family is considered, represented by the Pauli matrices (\ref{eq:sigmas}) compactified in the Minkowski metric.

\newpage
\begin{figure}
    \centering
    \includegraphics[width=1\linewidth]{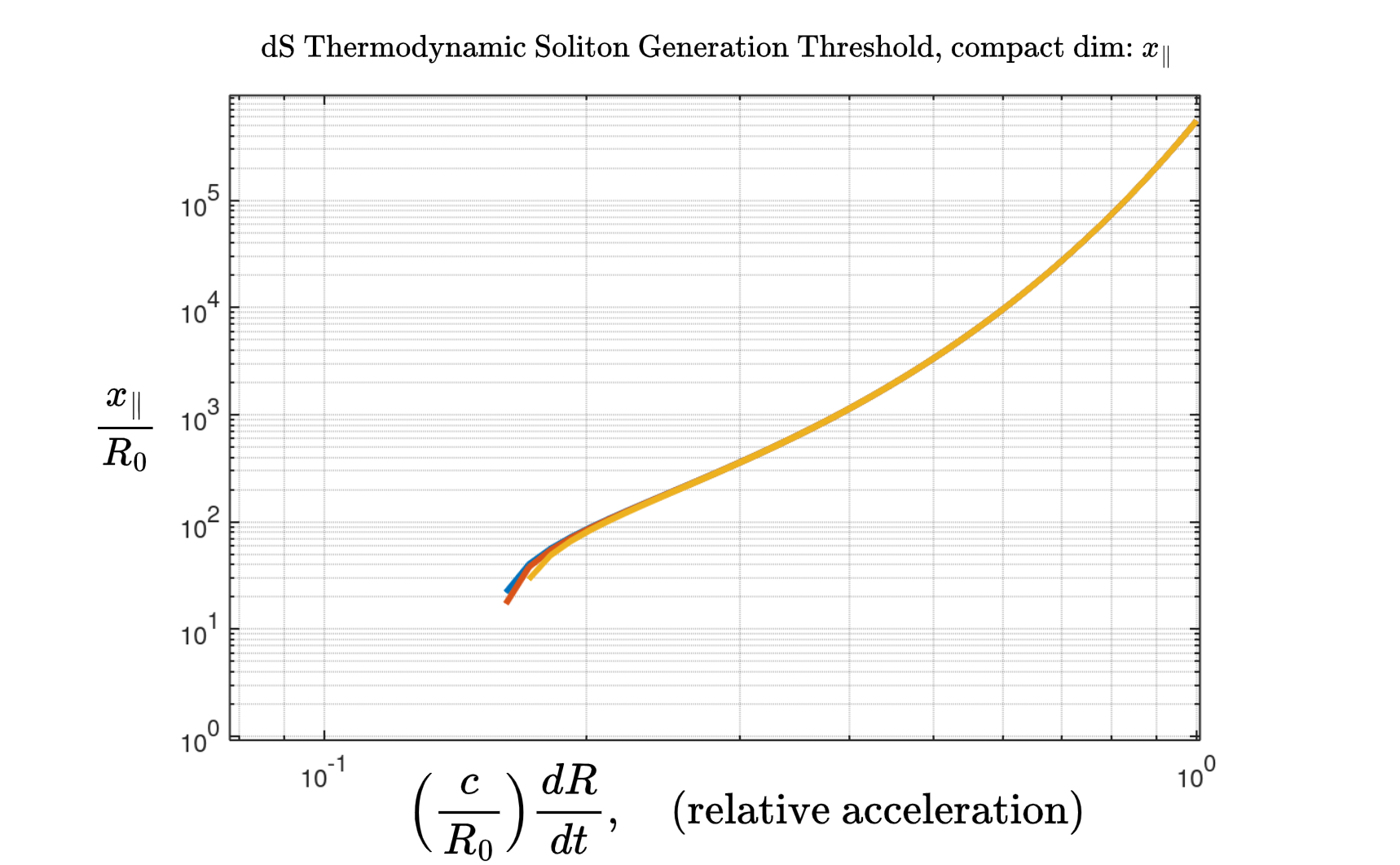}
    \caption{ Equating (\ref{eq:desitterpath}) with (\ref{eq:dslong}), $\frac{m}{r} = \frac{R}{r}= \frac{4}{(e-1)}$ for the soliton solution $\beta = \alpha = \infty$.  Perturbations $R_0$ (blue), $0.2R_0$ (red) and $0.1R_0$ (yellow). The deployment of the Kaluza-Klein compact dimension, $x_{\parallel}$ is subject to meeting the Machian thermodynamic threshold $\approx 0.15 \frac{c}{R_0} \frac{dR}{dt}$ for a topological defect to form. $R_0$ is the flux-quantized characteristic radius. For a superconducting electron ensemble $R_0$ is already large having a Meissner vortex radius $\sim 10^{-10}$m. A reverse energy cascade process such as a coherent electron pressure expansion to $2R_0$ requires sub-femtosecond time scale to generate a topological defect with $1$-$2$ nm radius. 
    An unconditioned, isotropic vacuum has $R$ from fig (\ref{fig:complete}) being related to the cosmological constant for de Sitter space.\cite{lochak_theory_2015} A suitably polarized fs laser must have a time constant of $0.22\frac{R_0}{c}$ based on an observability criteria of $\frac{x_{\parallel}}{R_0} \geq \hbar = \frac{\mu_0 c}{2\pi \sqrt{6}}$ as well as meeting power density requirements to completely polarize the vacuum volume with radius $R_{\text{final}}$. For a 1 fs pulse $R_{\text{final}} \sim 0.1 \mu$m to achieve topological defects with monopole radii, $r_M \geq \hbar$. }
    \label{fig:KK_deployment}
\end{figure}
\clearpage
\printbibliography
\end{document}